\documentclass{article}

\usepackage{xcolor}
\usepackage{color}

\usepackage{fullpage}

\usepackage{amsmath, amsthm}
\usepackage{amssymb}
\usepackage{graphicx}

\newtheorem{theorem}{Theorem}

\newcommand{\mioperatorname}[1]{\mathop{\mathrm{#1}}}

\newcommand{\loglog}{{\mioperatorname{ loglog}}}

\newcommand{\eqn}{\begin{eqnarray}}
\newcommand{\eqns}{\begin{eqnarray*}}
\newcommand{\uneq}{\end{eqnarray}}
\newcommand{\uneqs}{\end{eqnarray*}}

\newcommand{\ra}{\rightarrow}

\newcommand{\e}{\varepsilon}
\newcommand{\bits}{\{0,1\}}

\newcommand{\rr}{\mathbb{R}}
\newcommand{\zz}{\mathbb{Z}}

\newcommand{\ignore}[1]{}

\newcommand{\cc}{\mathbb{C}}

\newcommand{\fhat}{\hat{f}}
\newcommand{\ghat}{\hat{g}}

\author{John Steinberger, Tsinghua University}
\title{The sum-capture problem for abelian groups}

\begin{document}

\maketitle

\begin{abstract}
Let $G$ be a finite abelian group, let $0 < \alpha < 1$, and let $A
\subseteq G$ be a random set of size $|G|^\alpha$.  We let
$$
\mu(A) = \max_{B,C:|B|=|C|=|A|}|\{(a,b,c) \in A \times B \times C : a
= b + c \}|.
$$
The issue is to determine upper bounds on $\mu(A)$ that hold with high
probability over the random choice of $A$. Mennink and Preneel
\cite{BM} conjecture that $\mu(A)$ should be close to $|A|$ (up to
possible logarithmic factors in $|G|$) for $\alpha \leq 1/2$ and that
$\mu(A)$ should not much exceed $|A|^{3/2}$ for $\alpha \leq 2/3$.  We
prove the second half of this conjecture by showing that
$$
\mu(A) \leq |A|^3/|G| + 4|A|^{3/2}\ln(|G|)^{1/2}
$$
with high probability, for all $0 < \alpha < 1$. We note that $3\alpha
- 1 \leq (3/2)\alpha$ for $\alpha \leq 2/3$.

In previous work, Alon et al$.$ have shown that $\mu(A) \leq
O(1)|A|^3/|G|$ with high probability for $\alpha \geq 2/3$ while
Kiltz, Pietrzak and Szegedy show that $\mu(A) \leq |A|^{1 + 2\alpha}$
with high probability for $\alpha \leq 1/4$.  Current bounds on
$\mu(A)$ are essentially sharp for the range $2/3 \leq \alpha \leq
1$. Finding better bounds remains an open problem for the range $0 <
\alpha < 2/3$ and especially for the range $1/4 < \alpha < 2/3$ in
which the bound of Kiltz et al$.$ doesn't improve on the bound given
in this paper (even if that bound applied).  Moreover the conjecture
of Mennink and Preneel for $\alpha \leq 1/2$ remains open.
\end{abstract}

\section{Introduction}

Let $G$ be a finite abelian group, 
let $0 < \alpha < 1$, and
let 
$A \subseteq G$ be a random set of set of size $|G|^{\alpha}$. 
Define
$$
\mu(A) = 
\max_{B,C:|B|=|C|=|A|}|\{(a,b,c) \in A \times B \times C : a = b + c\}|.
$$
The main question we consider is to determine upper bounds on $\mu(A)$ that hold with high
probability over the random choice of $A$. 
We are motivated in particular by a 
conjecture of Preneel and Mennink
\cite{BM}, who posit the existence of constants $C_1$, $C_2$ such that
$$
\Pr[\mu(A) \geq C_1 |A| \log(|G|)] = o(1)
$$
for $\alpha \leq 1/2$ 
and such that
$$
\Pr[\mu(A) \geq C_2 |A|^{3/2}] = o(1)
$$
for $\alpha \leq 2/3$. 
We view $|G|$ as going to infinity, without further structural
assumptions on $G$. (The nature of the
abelian group, indeed, seems to have little influence\footnote{But 
  these conjectures are originally stated for
  $G = \zz_2^n$ in \cite{BM}.}.) 

Our main result is essentially 
to prove the second of the two conjectures
above. More precisely we show that
\eqn
\label{bbb} \Pr_A\left[\mu(A) \geq |A|^3/|G| + 4|A|^{3/2}\ln(|G|)^{1/2}\right]
\uneq
is negligible as $|G| \ra \infty$. 
Note the first term, $|A|^3/|G|$, is the expected size of the set
$$
\{(a,b,c) \in A \times B \times C : a = b + c\}
$$
when $A$, $B$ and $C$ are chosen at random. 
This term dominates for $\alpha > 2/3$
whereas the second term, $|A|^{3/2}\log(|G|)^{1/2}$, 
dominates for $\alpha < 2/3$.

More generally, if one defines
$$
\mu(A,B,C) = |\{(a,b,c) \in A \times B \times C: a = b + c\}|
$$
then we prove that
\eqn
\label{bibo} \Pr_A\left[\exists B, C \subseteq G\textrm{ s.t. } \mu(A,B,C) \geq |A||B||C|/|G| + 4\sqrt{\ln(|G|)|A||B||C|}\right]
\uneq
is negligible under the same assumptions as before (i.e$.$ that $|A|$
a fixed power of $|G|$ and that $|G|
\ra \infty$). The fact that \eqref{bbb} is negligible 
obviously a direct corollary of the fact that \eqref{bibo} is
negligible. 

We note these results can be given an interpretation in terms of
random Cayley graphs. More precisely, let $H_A$ be the Cayley graph of
vertex set $G$ and edge set associated to $A$, i.e., such that a
directed edge exists from $g_1 \in V(H_A)$ to $g_2 \in V(H_A)$ if and
only if $g_2 - g_1 \in A$. Then $\mu(A, B, C)$ is the number of edges
$(u,v)$ such that $v \in B \subseteq V(H_A)$ and $u \in -C
\subseteq V(H_A)$. Thus our result can loosely be paraphrased as:
with high probability over the choice of $A$ (with some fixed size),
the size of the largest subgraph induced by two shores of given sizes
is not much larger (in some sense) than if those shores were also chosen at random.

We note that
$$
|A||B||C|/|G| \geq \sqrt{|A||B||C|} \iff |A||B||C| \geq |G|^2
$$
so that \eqref{bibo} gives an essentially optimal pseudorandomness
result as long as $|A||B||C| \geq |G|^2$. 

In previous work (\cite{Alon} Theorem 4)  Alon et al$.$ show that
for every $0 < \alpha < 1$, $0 < \beta < 1$ such that $2\alpha + \beta
> 2 + 1/\loglog(|G|)$, 
\eqn
\label{alonbound} \Pr_A\left[\exists B \subseteq G, |B| \geq |G|^{\beta}\textrm{ s.t. }
  \mu(A,B,B) \geq \frac{\Theta(1)}{2\alpha + \beta - 2}|A||B|^2/|G| \right]
\uneq
is negligible, where $\Theta(1)$ denotes some absolute constant, and
where $A$ is again chosen uniformly at random from all subsets of $G$
of size $|G|^\alpha$. 
By comparison, \eqref{bibo} implies that for every $0
< \alpha < 1$, $0 < \beta < 1$ such that $\alpha + 2\beta > 2$ (i.e.,
such that $|A||B|^2 > |G|^2$),
$$
\Pr_A\left[\exists B \subseteq G, |B| \geq |G|^{\beta}\textrm{ s.t. }
  \mu(A,B,B) \geq (1 + c) |A||B|^2/|G| \right]
$$
is negligible for any constant $c > 0$. 
(This follows from the fact that $c|A||B|^2/|G| \geq
\sqrt{|A||B||B|}\log(|G|)^{1/2}$ 
when $\alpha + 2\beta > 2$.) 
Loosely speaking, thus, Alon
et al$.$ give an optimal pseudorandomness bound for 
$$
\max_B \mu(A, B, B)
$$
in the regime $|A|^2|B| \geq |G|^2$ whereas we give an optimal
pseudorandomness bound for the same quantity in the regime $|A||B|^2
\geq |G|^2$. The two bounds meet at $|A| = |B| = |G|^{2/3}$.

We also 
note that \eqref{alonbound} implies bounds on $\mu(A)$ for $\alpha >
2/3$. Namely, \eqref{alonbound} implies that for all $\alpha > 2/3$ there
exists a constant $c_\alpha = O(1/(3\alpha - 2))$ such that
$$
\Pr_A\left[\mu(A) \geq c_\alpha|A|^3/|G| \right]
$$
is negligible. This result, however, is superseded by
our observation that \eqref{bbb} is negligible. (Indeed, the latter implies
that $c_\alpha$ can in fact be taken any constant
greater than 1, independently 
of $\alpha$, and moreover supports $\alpha = 2/3$.)

In other, more recent related work, Kiltz et al$.$ \cite{kiltz} show that
$$
\Pr_A\left[\mu(A) \geq |A|^{1 + 2\alpha}\right]
$$
is negligible for all $0 < \alpha \leq 1/4$, where again $|A| =
|G|^\alpha$ and $\alpha$ is fixed as $|G|$ grows. This result 
shows
in particular  
that the exponent $3/2$ from \eqref{bbb} can be improved (and indeed
made arbitrarily close to 1) when $|A|$ is a small power of
$|G|$. Interestingly, $1 + 2\alpha = 3/2$ precisely when $\alpha =
1/4$, so our result 
implies the restriction $0 < \alpha \leq 1/4$ can be lifted while
essentially 
keeping the same bound. (In fact, while keeping a better bound, since
$3/2 < 1 + 2\alpha$ for $\alpha > 1/4$.) To summarize, the bound of
Kiltz et al$.$ on $\mu(A)$ is the best known for $0 < \alpha \leq 1/4$
while ours is the current state of the art for $1/4 < \alpha \leq
1$, and sharp bounds are only known for $2/3 \leq \alpha \leq 1$ (as
given variously by Alon et al$.$'s or by this paper).

It seems natural to conjecture that 
\eqn
\label{ddd} \mu(A) \approx \max(|A|, |A|^3/|G|)
\uneq
with high probability, up lower-order (e.g$.$, polylog$(|G|)$)
factors. If true, this would in particular imply that $\mu(A) \approx
|A|$ for $|A| \leq |G|^{1/2}$, as
conjectured by Mennink and Preneel \cite{BM}.
So far, however, \eqref{ddd} has only been established for $2/3 \leq
\alpha \leq 1$. \\

\noindent
\textsc{Techniques.} As might be guessed from the uncomplicated form of our
bound, our proof is very simple and uses only on basic discrete
Fourier analysis. More precisely, we rely on the fact that
$$
\mu(A, B, C) = \langle 1_A, 1_B * 1_C\rangle
$$
where $1_Z$ is the characteristic function of $Z \subseteq G$, where
$$
\langle f, g \rangle = \sum_{x \in G} f(x)g(x)
$$
is the inner product of two functions $f, g : G \ra \cc$, and where
$f * g$ is the convolution of functions $f$ and $g$, i.e.,
$$
(f * g)(x) = \sum_{y \in G} f(y) g(x-y)
$$
for all $x \in G$. We then use the fact that
\eqns
\langle 1_A, 1_B * 1_C\rangle & = & |G| \sum_S
\widehat{1_A}(S)\widehat{1_B *1_C}(S)\\
& = & |G|^2 \sum_S \widehat{1_A}(S) \widehat{1_B}(S)\widehat{1_C}(S)
\uneqs
where the sum is taken over the characters of $G$ and where $\hat{f}$
is the (discrete) Fourier transform of $f$. The fact that $A$ is
random is used to show that, with high probability, $|\widehat{1_A}(S)| \leq
4\sqrt{|A|\ln(|G|)}/|G|$ for all nontrivial characters $S$, where we
borrow the constant 4 from Hayes \cite{hayes}.
After
applying this observation, the result easily follows by separating the trivial
character from the rest of the sum, and by an application of
Cauchy-Schwarz.\\

\noindent
\textsc{Extensions.} Our main result and its corollaries also hold if
$A$ consists of $|G|^\alpha$ elements sampled uniformly at random 
with replacement. Indeed, this follows by inspection of the
proof of the afore-mentioned result of Hayes (\cite{hayes}, Lemma 6.3).\\

\noindent
\textsc{Applications.} We note that our result implies that the
compression function ``$\mathsf{F}_2$'' from \cite{BM} provably
achieves preimage resistance of $\sim 2^{2n/3}$ queries. Thus, of the preimage and collision resistance results in
\cite{BM}, only the collision-resistance of $\mathsf{F}_2$ remains
conjecture-based. \\

\noindent
\textsc{Acknowledgments.} We would like to thank Jooyoung Lee for
(re-)bringing
this problem to our attention, as well as Izabella Laba, who
suggested an earlier version of the title, and Mario Szegedy, who
suggested the current title. (Mario wins.)\\

\noindent
\textsc{Version History and a Missing Reference.} Shortly after posting this note on the
arxiv, J\'ozsef Solymosi, editor at E-JC, pointed out to us that very similar results are
already obtained in course notes of Babai \cite{babai}, a reference
which we (as well as the above-mentioned authors, except for Hayes)
had overlooked.
As our methods are basically the same as those of \cite{babai} 
this note thus offers basically nothing new. 
Our only merit is editorial: to bring the three 
groups of references \cite{babai, hayes}, \cite{Alon, kiltz}
and \cite{BM} to the attention of one another, as well as to give a
unified discussion of these previous results. While we have left the
rest of the paper untouched from the first version, we make no
longer make any claims to originality.

\section{Proof}

Since part of the intended audience for this paper 
are symmetric-key cryptographers (indeed, \cite{BM, kiltz} are both
cryptography papers, and this result seems to have other likely
applications in symmetric-key cryptography security proofs\footnote{In
a nutshell, this seems to come about as follows: in many cryptographic
security
proofs an adversary makes ``queries'' whose answers are randomly drawn
from a group $G$, e.g$.$ $G = \zz_2^n$; these queries form the set
$A$. One must then show that with
high probability these queries contain no unexpectedly ``helpful
structure'' for the adversary. The ``helpful structure'' might be,
in certain cases, a high value of $\mu(A)$.} \cite{seurin}) 
whose Fourier analysis might be a bit rusty, and since anyway the
proof is quite short, we take the leisure of developing the required Fourier
analysis from scratch. For notational convenience we assume that $G =
\zz_2^n$. Adapting the argument to an arbitrary group is
straightforward (this will be evident from the proof).

Let $G = \zz_2^n$. We identify $G$ with the set $\bits^n$ of binary
strings of length $n$. For $S \subseteq [n] = \{1, \ldots, n\}$ we recall that the
character function
function $\chi_S : \bits^n \ra \{-1,1\}$ is defined by 
$$
\chi_S(x) = \prod_{i\in S}(-1)^{x_i} = (-1)^{\sum_{i\in S}x_i}
$$
where $x = (x_1, \ldots, x_n) \in G = \bits^n$. 
Then $\chi_\phi = 1$ and $\chi_S$, $\chi_T$ are orthogonal for all $S
\ne T$, i.e.,
$$
\sum_{x\in \bits^n}\chi_S(x)\chi_T(x) = 0.
$$
Thus, also,
$$
E[\chi_S\chi_T] = 0,\qquad S \ne T
$$
where $E[f]$ is a shorthand for
$$
E_x[f(x)] = \frac{1}{|G|}\sum_{x\in\bits^n}f(x).
$$
More precisely,
$$
E[\chi_S\chi_T] = \begin{cases} 1 & \textrm{if $S = T$},\\ 0 & \textrm{if
    $S \ne T$.}\end{cases}
$$
Since $\chi_S\chi_T = \chi_{S\triangle T}$ (where $S \triangle T$ is
the summetric difference of $S$ and $T$), we note 
this reduces to the fact that
$$
E[\chi_S] = \begin{cases} 1 & \textrm{if $S = \phi$}, \\ 0 & \textrm{if
    $S \ne \phi$.}\end{cases}
$$

Every function $f : \bits^n \ra \rr$ can be seen as an element of
$\rr^{|G|}$. 
Since $\{\chi_S : S \subseteq [n]\}$ is a set of $|G|$ orthogonal
functions in $\rr^{|G|}$, they form a basis of $\rr^{|G|}$. I.e., for
every function $f : \bits^n \ra \rr$ there exist real numbers
$\alpha_S, S \subseteq [n]$ such that
$$
f = \sum_{S \subseteq [n]} \alpha_S \chi_S.
$$
The coefficients $\alpha_S$ are called the \emph{fourier coefficients}
of $f$ and are typically written $\fhat(S) := \alpha_S$. Thus
$$
f = \sum_{S \subseteq [n]} \fhat(S) \chi_S
$$
for any $f : \bits^n \ra \rr$. One has
$$
\fhat(S) = E[f\chi_S].
$$
More precisely, this can be verified from the fact that
$$
E[f\chi_S] = E\left[\left(\sum_{T \subseteq [n]} \alpha_T \chi_T\right)
  \chi_S\right] = E[\alpha_S\chi_S\chi_S] = \alpha_S
$$
using orthogonality.

We have
$$
E[fg] = E\left[\left(\sum_{T \subseteq [n]} \fhat(T) \chi_T\right)
\left(\sum_{S \subseteq [n]} \ghat(S) \chi_S\right)\right] =
E[\sum_{S\subseteq [n]}\fhat(S)\ghat(S)] = \sum_{S\subseteq [n]} \fhat(S)\ghat(S).
$$
for any $f, g : \bits^n \ra \rr$. In particular
$$
E[f^2] = \sum_{S\subseteq[n]} \fhat(S)^2
$$
and if $f : \bits^n \ra \{-1, 1\}$ then
$$
\sum_{S \subseteq [n]} \fhat(S)^2 = 1
$$
since $E[f^2] = 1$.

Moreover if $f : \bits^n \ra \{0,1\}$ then $(-1)^f : \bits^n \ra
\{-1,1\}$ and $(-1)^f = 1 - 2f$ so
\eqns
1 & = & \sum_{S\subseteq[n]}\widehat{(-1)^f}(S)^2\\
& = & \sum_{S\subseteq[n]}\widehat{1 - 2f}(S)^2\\
& = & \sum_{S\subseteq[n]} (\hat{1}(S) - 2\fhat(S))^2\\
& = & \sum_{S\subseteq[n]} \hat{1}(S)^2 - 4\hat{1}(S)\fhat(S) + 4\fhat(S)^2\\
& = & 1 - 4\fhat(\phi) + 4\sum_{S \subseteq [n]}\fhat(S)^2
\uneqs
from which we deduce:
$$
\fhat(\phi) = \sum_{S\subseteq[n]} \fhat(S)^2,\qquad\qquad(f : \bits^n
\ra \bits).
$$

Define
$$
(f*g)(x) = \sum_{y\in\bits^n} f(y)g(x+y) = |G| E_y[f(y)g(x+y)].
$$
(Note $x + y = x - y$ in the current group.) Using the fact that
$\chi_S(x + y) = \chi_S(x)\chi_S(y)$ for all $S$, $x$, $y$ we find
\eqns
\widehat{f*g}(S) & = & E_x[(f*g)(x)\chi_S(x)]\\
& = & E_x\left[\sum_y f(y)g(x + y)\chi_S(x)\right]\\
& = & |G|\sum_y f(y) \sum_x g(x+y)\chi_S(x)\\
& = & |G|\sum_y f(y) \sum_x g(x) \chi_S(x + y)\\
& = & |G|\left(\sum_y f(y)\chi_S(y)\right)\left(\sum_x
  g(x)\chi_S(x)\right)\\
& = & |G| \fhat(S) \ghat(S).
\uneqs

We write $1_Z$ for the characteristic function of a set $Z \subseteq
\bits^n$.
Note that for sets $A, B, C \subseteq \bits^n$ we have 
\eqns
|\{(z, a, b) \in A \times B \times C : z = a + b\}| 
&  = & 
\sum_{x\in\bits^n}1_A(x)(1_B * 1_C)(x) \\
& = & |G| E[1_A(1_B * 1_C)]\\
& = & |G|\sum_{S\subseteq[n]} \widehat{1_A}(S)\widehat{1_B*1_C}(S)\\
& = & |G|^2\sum_{S\subseteq[n]}
\widehat{1_A}(S)\widehat{1_B}(S)\widehat{1_C}(S)
\uneqs

\newcommand{\kappabar}{\overline{\kappa}}

Now let $A \subseteq \bits^n$ consist of $|G|^\alpha$ elements
sampled uniformly at random without replacement, 
Fix $S \subseteq [n]$, $S \ne \emptyset$. Let $\chi_S^+ = \{x \in
\bits^n : \chi_S(x) = 1\}$, $\chi_S^- = \{x \in \bits^n : \chi_S(x) =
-1\}$ be the supports of the positive and negative supports of
$\chi_S$. Note $|\chi_S^+| = |\chi_S^-| = |G|/2$ and that
$$
|G|\cdot \widehat{1_A}(S) = |A\cap \chi_S^+| - |A\cap \chi_S^-|.
$$
Since the points in $A$ are uniformly distributed in $\bits^n$,
$|G|\cdot \widehat{1_A}(S)$ is therefore concentrated around 0. 
If $A$ were sampled uniformly with replacement, a Chernoff bound
would show
$$
\Pr\left[|G|\cdot |\widehat{1_A}(S)| \geq c\sqrt{|A|} \right] \leq 2e^{-c^2/2}.
$$
which would imply that, with high probability over the choice of $A$,
$$
|\widehat{1_A}(S)| \leq \frac{1}{|G|}\sqrt{(2+h)\ln(|G|)|A|}
$$
for all $S \ne \emptyset$, where $h > 0$ can be any fixed value.
Unfortunately $A$ is sampled without replacement so Chernoff bounds
must be eschewed in favor of Martingales and of Azuma-type inequalities. 
Such results, in fact, have already been obtained by Hayes
\cite{hayes},
who among others proves the following:

\begin{theorem} [Hayes, \cite{hayes} Theorem 1.13]
Let $\e > 0$. Let G be a finite abelian group, and let $0 \leq m
\leq |G|$. 
For all but an $O(|G|^{-\e})$ fraction of subsets $A \subseteq G$ such
that $|A| = m$, the maximum non-principal fourier coefficient of $1_A$ is
upper bounded by 
$$
\frac{2}{|G|}\sqrt{2(1 + \e) \ln(|G|)m'}
$$
in absolute value, where $m' = \min(m, |G| - m)$.
\end{theorem}

\noindent
In particular, returning to $G = \zz_2^n$ (although
this choice of $G$ will play in an increasingly small role in the
remainder), and setting (say) $\e = 1$ is Hayes's theorem,
we have
\eqn
\label{bizu} |\widehat{1_A}(S)| \leq \frac{4}{|G|} \sqrt{\ln(|G|)|A|}
\uneq
for all $S \subseteq [n]$, $S \ne \phi$, with overwhelming probability
over the choice of $A$, $|A| = |G|^\alpha$. 
For what follows, we assume such a ``generic'' $A$.
Then for all $B, C \subseteq G$ we have
\eqns
|\{(a, b, c) \in A \times B \times C : a = b + c\}| 
& = & |G|^2\sum_{S\subseteq[n]}
\widehat{1_A}(S)\widehat{1_B}(S)\widehat{1_C}(S)\\
& = & |G|^2\left(\frac{|A|}{|G|}\frac{|B|}{|G|}\frac{|C|}{|G|} +
  \sum_{S\ne \phi}\widehat{1_A}(S)\widehat{1_B}(S)\widehat{1_C}(S)\right)\\
& \leq &  \frac{|A||B||C|}{|G|} + |G|^2\sum_{S\ne \phi}|\widehat{1_A}(S)|\widehat{1_B}(S)\widehat{1_C}(S).
\uneqs
Note that
$$
\sum_{S \ne \phi}
\widehat{1_B}(S)\widehat{1_C}(S) \leq \sqrt{\sum_{S\subseteq[n]}
  \widehat{1_B}(S)^2}\sqrt{\sum_{S\subseteq[n]} \widehat{1_C}(S)^2} =
\sqrt{\widehat{1_B}(\phi)}\sqrt{\widehat{1_C}(\phi)}
= \frac{1}{|G|}\sqrt{|B||C|}
$$
by Cauchy-Schwarz. So, by \eqref{bizu},
\eqn
\label{up1}
\sum_{S\ne \phi}
|\widehat{1_A}(S)| \widehat{1_B}(S)\widehat{1_C}(S) \leq \frac{4}{|G|^2}\sqrt{\ln(|G|)|A||B||C|}
\uneq
and, altogether,
\eqns
|\{(a, b, c) \in A \times B \times C : a = b + c\}| \leq
\frac{|A||B||C|}{|G|} + 4\sqrt{\ln(|G|)|A||B||C|}
\uneqs
for all sets $B, C \subseteq G$. (Looking back on the proof, we note
that the constant 4 can be replaced with $2\sqrt{2} + h$ for any $h >
0$.)

\end{document}